\documentclass[twocolumn]{aastex62}

\usepackage{color, mhchem}

\graphicspath{{./}{figures/}}

\accepted{\today}
\submitjournal{ApJ}

\shorttitle{$^{13}$CCH in the TW Hya Disk}
\shortauthors{Bergin et al.}

\begin{document}

\title{The Carbon Isotopic Ratio and Planet Formation}

\correspondingauthor{Edwin A. Bergin}
\email{ebergin@umich.edu}

\author{Edwin A. Bergin}
\author{Arthur Bosman}
\affil{Department of Astronomy,
University of Michigan, 1085 S. University Ave,  Ann Arbor, MI 48109, USA}

\author{Richard Teague}
\affil{Department of Earth, Atmospheric, and Planetary Sciences, Massachusetts Institute of Technology, Cambridge, MA 02139}

\author{Jenny Calahan}
\affil{Department of Astronomy,
University of Michigan, 1085 S. University Ave,  Ann Arbor, MI 48109, USA}

\author{Karen Willacy}
\affil{Jet Propulsion Laboratory, California Institute of Technology, 4800 Oak Grove Dr., Pasadena, CA 91109, USA}

\author{L. Ilsedore Cleeves}
\affil{Department of Astronomy, University of Virginia, Charlottesville, VA 22904}

\author{Kamber Schwarz}
\affil{Max Planck Institute for Astronomy, K\"{o}nigstuhl 17, Heidelberg, Germany}

\author{Ke Zhang}
\affil{Department of Astronomy, University of Wisconsin-Madison, 475 N. Charter St., Madison, WI 53706}

\author{Simon Bruderer}
\affil{Max-Planck-Institut f{\"u}r Extraterrestrische Physik, Giessenbachstrasse 1, 85748 Garching, Germany}

\begin{abstract}
We present the first detection of $^{13}$CCH in a protoplanetary disk (TW Hya).  Using observations of C$_2$H we measure \ce{CCH}/\ce{$^{13}$CCH} = 65$\pm$20 gas with a CO isotopic ratio of $^{12}$CO/$^{13}$CO = 21$\pm$5 \citep{Yoshida22}.   The TW Hya disk exhibits a gas phase C/O that exceeds unity and \ce{C2H} is the tracer of this excess carbon.  We confirm that the TW Hya gaseous disk exhibits two separate carbon isotopic reservoirs as noted previously \citep{Yoshida22}.   We explore two theoretical solutions for the development of this dichotomy.  One model represents TW Hya today with a protoplanetary disk exposed to a cosmic ray ionization rate that is below interstellar as consistent with current estimates.  We find that this model does not have sufficient ionization in cold (T $<$ 40~K) layers to activate carbon isotopic fractionation.   The second model investigates a younger TW Hya protostellar disk exposed to an interstellar cosmic ray ionization rate.  We find that the younger model has sources of ionization deeper in a colder disk that generates two independent isotopic reservoirs.  One reservoir is $^{12}$C-enriched carried by methane/hydrocarbon ices and the other is $^{13}$C-enriched carried by gaseous CO.  The former potentially provides a source of methane/hydrocarbon ices to power the chemistry that generates the anomalously strong C$_2$H emission in this (and other) disk systems in later stages. The latter provides a source of gaseous $^{13}$C rich material to generate isotopic enrichments in forming giant planets as recently detected in the  super-Jupiter TYC 8998-760-1 b by \citet{Zhang21_13coexo}.
\end{abstract}


\def\twCO{$^{12}$CO}           
\def\thCO{$^{13}$CO}           
\def\RCO{$^{12}$CO/$^{13}$CO}  
\def\RC{$^{12}$C/$^{13}$C}  
\def\twCp{$^{12}$C$^+$}           
\def\thCp{$^{13}$C$^+$}           

\section{Introduction} \label{sec:intro}

The exoplanet revolution is now heading in a phase of greater characterization where the composition of planetary atmospheres can be explored in greater detail via high resolution spectroscopy \citep[e.g.,][]{Line21, Zhang21_13coexo, Brogi_Birkby21} or observations with new more sensitive instruments \citep[e.g., JWST;][]{JWST_co2}.  A greater understanding of composition allows for an exploration of how the formation conditions of the planet might be reflected in its composition.  Over the past decade the strongest link has been to explore the bulk elemental abundance ratio of carbon to oxygen or C/O.  \citet{omb11} first presented this idea utilizing basic chemistry (e.g., deposition/sublimation of primary molecular carriers of C and O) to show that the gas phase and solid-state C/O ratios are predicted to change with distance from the star and it might be possible to tie composition to the formation location.
 This may be more complicated as planets migrate, can accrete icy planetesimals/pebbles, and might have core-atmosphere mixing along with gravitational settling \citep{Cridland16,Helled17, Guillot2022}. However, it captures a central element and is now widely compared to exoplanet atmospheric composition retrievals \citep[][to list a few]{Barman15, Lavie17, Oreshenko17}.

Beyond the C/O ratio, the carbon isotopic ratio presents another interesting avenue for exploring the links between disk and exoplanet composition.   This has been enabled by advances in high resolution spectroscopy of exoplanets that have isolated the $^{12}$C/$^{13}$C ratio in two systems.  In one case for a young, accreting super-Jupiter TYC 8998-760-1 b at 160 au, \citet{Zhang21_13coexo}, measure a ratio of $^{12}$C/$^{13}$C = 31$^{+17}_{-10}$ (90\% confidence).  This is a significant $^{13}$C enrichment when compared to the local interstellar value of 68 \citep{Langer93, Milam05f}.  A similar level of enrichment is found towards the Hot Jupiter WASP-77Ab, $^{12}$C/$^{13}$C = 10.2-42.6 at 68\% confidence \citep{Line21}.  This enrichment is posited to have originated in molecular ices \citep{Zhang21_13coexo}.
Curiously, the solar system betrays no evidence of $^{13}$C enrichments at this level in any body, including comets that are comprised of molecular ices \citep{Nomura22}.   Further, interstellar ices towards quiescent lines of sight also show little evidence of enrichment \citep{McClure23}.   This points towards later phases, i.e., disk chemical evolution.

If molecular ices are enriched in $^{13}$C within the disk, we should see some evidence of this in disk systems as the gas should exhibit the opposite (i.e., a depletion), provided the fractionation originates in the disk. The state-of-the-art of these measurements can be found in TW Hya, the nearest young gas-rich disk, and the best characterized system.  
\citet{Yoshida22} used the line wing emission of optically thick CO isotopologues to derive the carbon isotopic ratio in {\em gaseous} CO as a function of position.  They find that $^{13}$C isotopic enrichments are present in disk surface layers from $\sim$70-100 au. The detection of $^{13}$C$^{18}$O inside the CO snowline implies $^{12}$CO/$^{13}$CO $\sim 40^{+9}_{-6}$ in sublimated CO ices \citep{Zhang17}. In contrast,  the outer ($>$100~au) disk shows the opposite signature and appears to be depleted in $^{13}$CO \citep{Yoshida22}.  

However, the C/O ratio in TW Hya is measured to be $>$ 1 \citep{Bergin16, Kama16a}.  Thus, CO is not the only reservoir of carbon.   
Here we report the detection of weak $^{13}$CCH emission within the TW Hya disk.  The chemistry of CCH is the tracer of the excess carbon not carried by CO providing accurate measure of its isotopic ratio.   
 To explore the origin of isotopic enhancements considering the full extent of isotopic measurements we use models of carbon fractionation of TW Hya today (a later stage few Myr disk, i.e. class II) and in an early-stage (i.e. less evolved; class I) state.      In \S 2 we will present our observations and the technique used to obtain the weak signal.  \S 3 presents the derived isotopic ratio and the chemical models that are used to explore fractionation in the disk environment.  Finally, in \S 4 we discuss  the implications of this result and our conclusions.

\section{Observations}

\subsection{\rm C$_2$H}

The C$_2$H data we use was originally published in \citet{Bergin16} using the data from ALMA project 2013.1.00198.S. We refer the reader to the original article for details on the data reduction and imaging. This included the subtraction of the continuum by modeling the continuum contribution as a linear component around the line emission and using the \texttt{uvcontsub} task to remove this in the image plane. Note that as the continuum emission is confined to inside of ${\sim}~1\arcsec$ (${\sim}~60$~au), the continuum subtraction should have minimal effect on the line data presented here. The resulting images have a beam of $0.50\arcsec \times 0.41\arcsec$ with a position angle of $59.0\degr$.

\begin{figure}
    \centering
    \includegraphics[width=0.9\columnwidth]{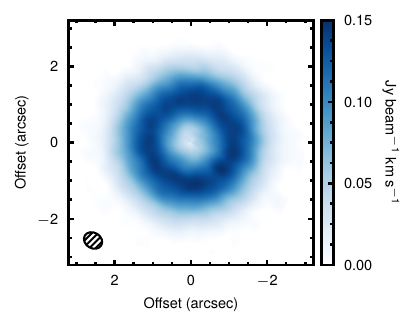}
    \caption{Zeroth moment of the $N = 4-3$ $J = 9/2 - 7/2$ $F = 5-4$ transition of C$_2$H made using a Keplerian mask. The peak value is $148.7~{\rm mJy~beam^{-1}~km\,s^{-1}}$. The beam size is shown in the bottom left of the panel.}
    \label{fig:CCH_zeroth}
\end{figure}

Figure~\ref{fig:CCH_zeroth} shows the integrated flux density for the $N = 4-3$ $J = 9/2 - 7/2$ $F = 5-4$ transition made using a Keplerian mask assuming $i = 6.8\degr$, ${\rm PA} = 151\degr$, $M_{\rm star} = 0.6~M_{\rm sun}$ and $d = 60.1~{\rm pc}$. The mask was used rather than $\sigma$-clipping to remove any contribution from the $F = 4-3$ transition. The radial profile is well described by a Gaussian centered at $1.1\arcsec$ with a width of $0.58\arcsec$ \citep{Bergin16}.

Integrating  over the whole disk yields a total integrated flux of $6.17 \pm 1.16~{\rm Jy~km\,s^{-1}}$ where the uncertainties have been calculated following \citet{Teague19}, while over the annulus with an inner radius of $0.52\arcsec$ (31~au) and an out radius of $1.68\arcsec$ (101~au) we find $4.20 \pm 0.68~{\rm Jy~km\,s^{-1}}$. The peak intensity was $372 \pm 6~{\rm mJy\,beam^{-1}}$.\footnote{There is a difference in this intensity and that given in Table~1 of \citet{Bergin16} where the peak intensity for this transition is listed as $1.100 \pm 0.007~{\rm Jy~beam^{-1}}$.  In this work we deblended the F = 5--4 and 4--3 which was not done by \citet{Bergin16}.}

The non-Gaussian beam which results from the imaging process can introduce subtle inaccuracies when calculated integrated fluxes. Following \citet{CZekala_ea_2021}, the non-Gaussianity of the beam can be quantified with $\epsilon$, which is the ratio of the volumes of the CLEAN and dirty beams, where $\epsilon = 1$ indicates a purely Gaussian beam. Using the scripts released as part of the MAPS Large Program \citep{Oberg21} to apply a correction to this (the so-called `JvM correction'), the C$_2$H data were found to have $\epsilon = 0.69$. Calculating the integrated flux density and peak intensity with the JvM-corrected data found discrepancies of 5\% and 2\%, respectively. These uncertainties were incorporated in the ratios described below by adding this uncertainty in quadrature to the statistical uncertainty.

\subsection{\rm $^{13}$CCH}

The $^{13}$CCH line observations of TW Hya were carried out by ALMA on November 23, 2016 (2015.1.00308.S). The array included 42 antennas, with baselines lengths spanning from 15 to 704\,m. The correlators were set to have three-line windows centered on 336.571, 351.060, and 349.346\,GHz, and each window had a bandwidth of 467\,MHz. A fourth spectral window was dedicated to continuum observations, centering on 338.769\,GHz with a band width of 2\,GHz. The Nearby Quasars J1037-2934, J1058+0133, J1107+4449 were used for gain, bandpass, and flux calibration, respectively.  The total on-source time was 49\,minutes. The raw visibility data were calibrated by NRAO staff members in CASA version 4.7.0, and we further applied phase-only self-calibration on the data using the continuum spectral window and line-free channels in all three-line spectral windows. The fully calibrated visibilities were then Fourier inverted to generate images by using the CLEAN task with a Briggs weight of 0.5. The final $^{13}$CCH image has a noise level of $3.6~{\rm mJy \, beam^{-1}}$ per $210~{\rm m\,s^{-1}}$ channel and a beam size of $0.32\arcsec \times 0.30\arcsec$ with a position angle of $4.4\degr$.

To extract a disk-averaged spectrum we use the Python package \texttt{GoFish} \citep{GoFish} which implements the method presented in \citet{Yen16}. First the spectrum in each pixel was shifted to the systemic velocity by applying a shift of $\delta v(r,\,\phi) = -v_{\rm kep}(r) \, \cos \phi \, \sin i$, where $v_{\rm kep}(r)$ is the Keplerian rotation at radius $r$, $\phi$ is the polar angle of the pixel, measured relative to the red-shifted major-axis, and $i$ is the disk inclination. With all spectra centered on the systemic velocity, they are stacked to increase the signal-to-noise of the spectrum. This method will result in the average spectrum in a pixel within the averaged area. To obtain an integrated flux density, this average spectrum is scaled by the total area of the annulus over which the spectra were averaged. This approach of first averaging and then scaling is mathematically the same as integrated over the entire area, however the intermediate averaged spectrum allows for a more accurate determination of the integration boundaries. 

For the velocity shifting we assume the same parameters as for the Keplerian mask in the previous section, $i = 6.8\degr$, ${\rm PA} = 151\degr$, $M_{\rm star} = 0.6~M_{\rm sun}$ and $d = 60.1~{\rm pc}$. Assuming that the radial emission profile of $^{13}$CCH follows that of C$_2$H, we average from $0.52\arcsec$ (31~au) to $1.68\arcsec$ (101~au) to increase the signal-to-noise ratio of the detection, weighting each individual annulus by their respective area. As each annulus has a fixed width of 1/4 of the beam FWHM (${\sim}~80~{\rm mas}$), the weights are simply proportional to $r^2$.

To verify that the image was centered, we vary the source center by small pixel-scale offsets and measure the signal-to-noise ratio of spectrum. We find a peak signal-to-noise with an offset of $(\delta x_0,\, \delta y_0) = (-52~{\rm mas}, -155~{\rm mas})$.

\begin{figure}
    \centering
    \includegraphics[width=\columnwidth]{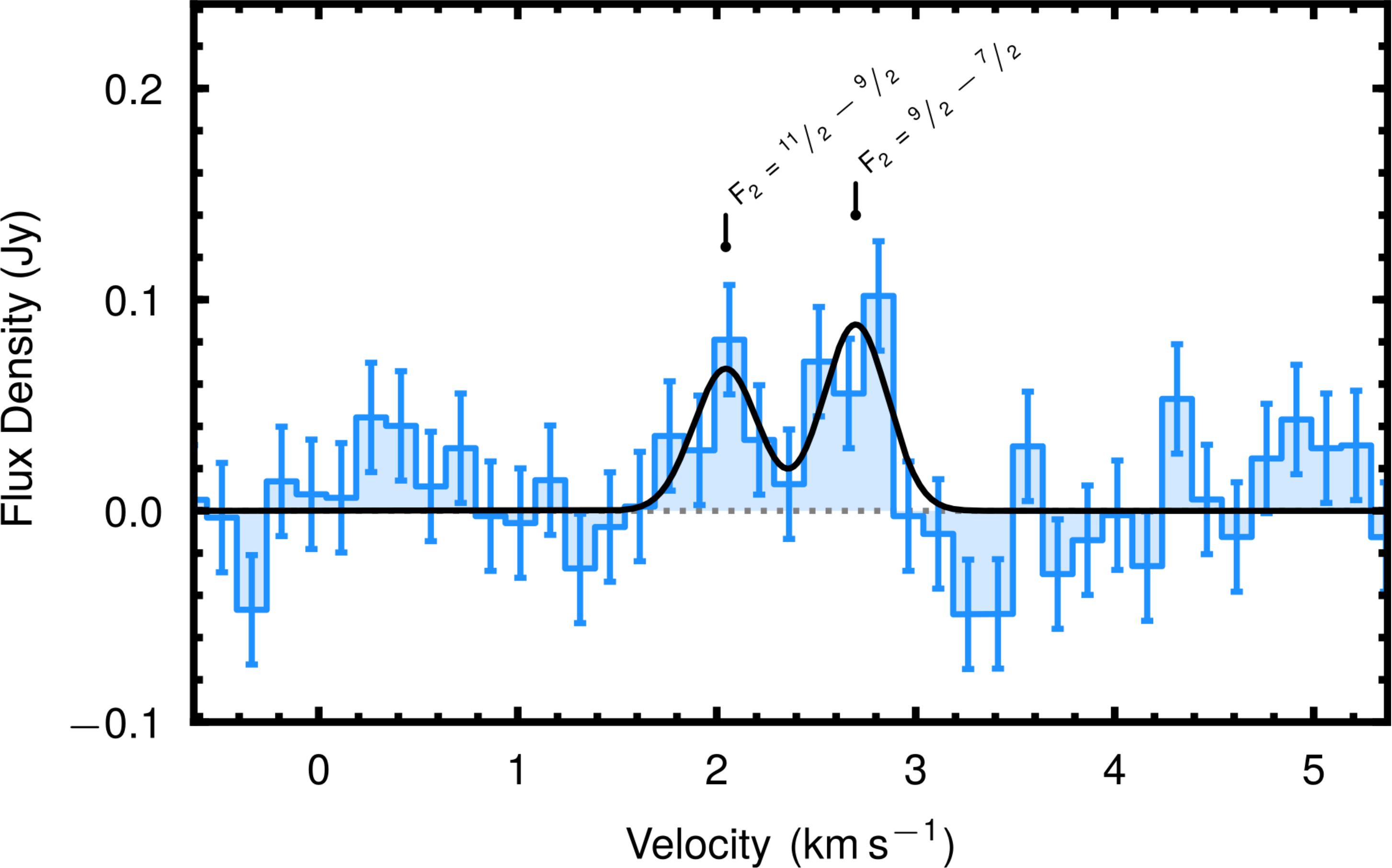}
    \caption{Integrated average flux density between 0.8\arcsec{} and 1.8\arcsec{}, resampled to a velocity resolution of 150~m\,s$^{-1}$, clearly showing the F = 11/2 - 9/2 and F = 9/2 - 7/2 components of the N = 4 - 3, J = 9/2 - 7/2 transition. The black lines show two Gaussian fits to the spectrum with a shared line width of 125~m\,s$^{-1}$.}
    \label{fig:13CCH_spectrum}
\end{figure}

The resulting spectrum is shown in  Fig.~\ref{fig:13CCH_spectrum}, showing the clear detection of the F$_2$ = 11/2 - 9/2 and F$_2$ = 9/2 - 7/2 components of the N = 4 - 3, J = 9/2 - 7/2 F$_1$ = 5 - 4 transition. Integrating this spectrum yields an integrated flux of $56.9 \pm 14.1~{\rm mJy~km\,s^{-1}}$. We have verified by changing the inner and outer radii of the annuli we integrate over that this range gives the highest signal-to-noise detection of the lines. Integrating over the whole disk finds an integrated flux of $70.2 \pm 23.2~{\rm mJy~km\,s^{-1}}$. From Gaussian fits to the line, the peak intensity values for the bright annulus region were $0.9 \pm 0.3~{\rm mJy~beam^{-1}}$ and $1.2 \pm 0.3~{\rm mJy~beam^{-1}}$ for the F$_2$ = 11/2 - 9/2 and F$_2$ = 9/2 - 7/2 components, respectively.

To verify that the non-Gaussianity of the beam did not bias the results, the JvM-correction \citep{CZekala_ea_2021} was applied to the $^{13}$CCH data. These data were found to have $\epsilon = 0.70$, consistent with the C$_2$H data, which is unsurprising given they use the same antenna configuration. The integrated flux density was not found to substantially vary compared to the native data, finding only a 1\% difference. The peak intensity varied more, up to 40\%, however this is because such a low-level signal is dominated by the noise which is substantially suppressed with the JvM-correction. As such, this term is ignored as the difference is equivalent to the difference in the noise levels determined by the RMS of a line-free channel. The 1\% uncertainty on the integrated flux density is added in quadrature to the statistical uncertainties.

\section{Carbon Isotopic Ratio in TW Hya}

\subsection{\rm $^{12}$C$^{12}$H/$^{13}$C$^{12}$H in TW Hya}

Using the integrated intensities from above, we find a C$_2$H / $^{13}$CCH ratio of $74 \pm 22$ in the bright annulus and $88 \pm 3\mathbf{4}$ averaged over the whole disk. Similarly, if we use the peak flux density for each line, after correcting for the relative beam sizes and assuming that they're optically thin so that the two F$_2$ components can be added, we find a C$_2$H / $^{13}$CCH ratio of $78 \pm 15$, consistent with the integrated flux approach.   The above ratios are ratios in flux and not total column.  Based on the fact that the flux ratio is consistent with the interstellar $^{12}$C/$^{13}$C isotopic ratio \citep[$\sim$68][]{Langer93, Milam05f} and hyperfine ratios \citep{Bergin16}, the C$_2$H emission in TW Hya is optically thin.   In this limit, for observations that have been corrected for differences in the beam size (as above), the integrated flux density ratio is essentially equivalent to the overall column density ratio correcting for an 11\% difference in the A-coefficents. We therefore conclude that N(CCH)/N($^{13}$CCH) = 65$\pm$20 over spatial scales commensurate with the TW Hya C$_2$H ring.

\begin{figure*}
    \centering
    \includegraphics[width=0.8\textwidth]{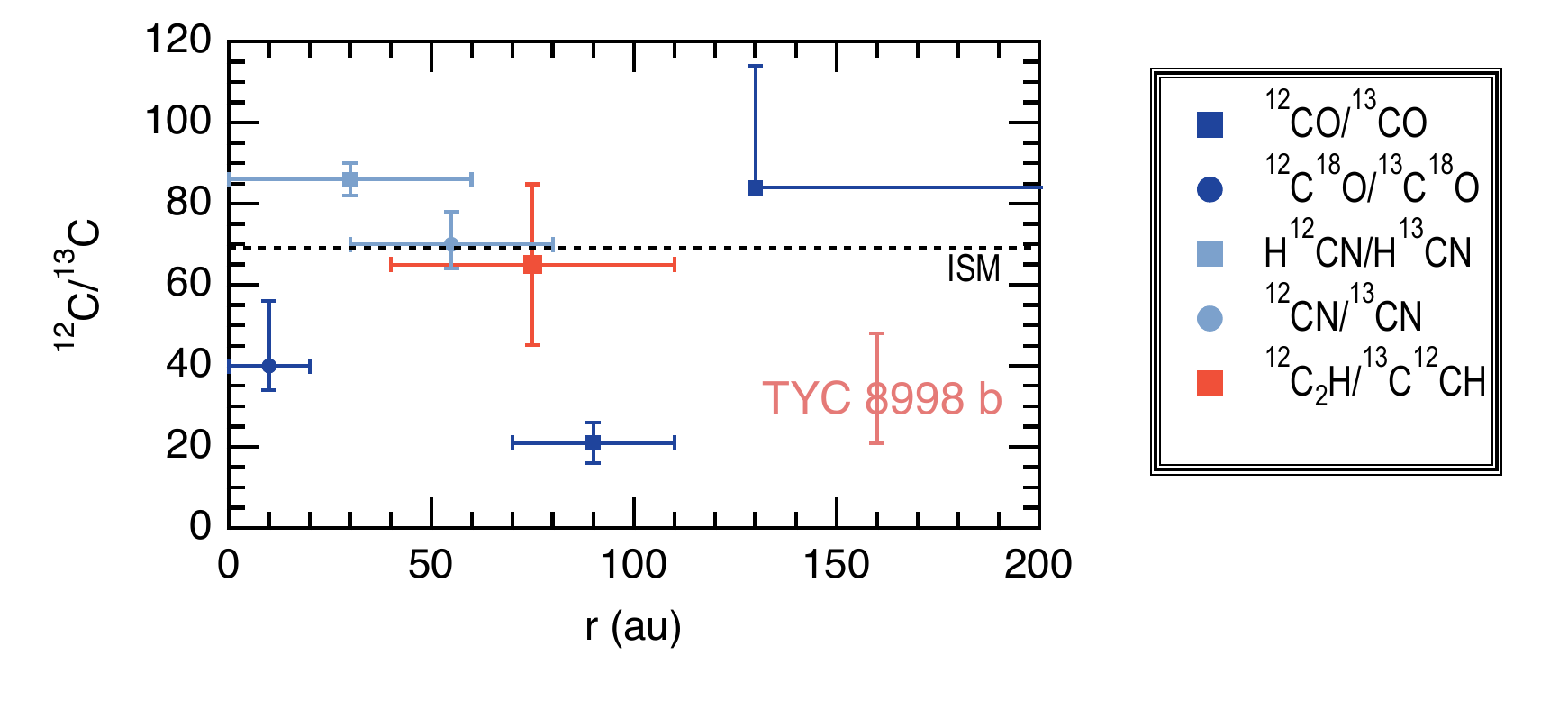}
    \caption{Measurements of isotopic ratios in TW Hya placed into observational context in terms of distance from the star.  Measurements are from \citet[][CO]{Yoshida22}, \citet[][C$^{18}$O]{Zhang17}, \citet[][HCN]{Hily-Blant19}, and CCH (presented here).  For illustrative purposes, we also include the measurement of TYC 8998 b \citep{Zhang21_13coexo} which is shown at the orbital distance estimated from its host star.}
    \label{fig:isointwhya}
\end{figure*}

\bigskip
\bigskip
\subsection{Two Carbon Isotope Reservoirs}

Fig.~\ref{fig:isointwhya} provides the landscape of isotopic ratio measurements within TW Hya, with explicit values given in Table~\ref{tab:tw_iso}.   Inside of the CO snowline at 21~au \citep{Schwarz16}  an isotopic ratio of $\sim$40$^{+9}_{-6}$ is measured via detection of \ce{^{13}C^{18}O} by \citet{Zhang17}.  \citet{Hily-Blant19} observe HCN and \ce{H^{13}CN} and measure an average ratio of 86$\pm4$ inside 60 au.  The differences between CO, HCN, and CN provide evidence of multiple isotopic reservoirs in the system as noted by \citet{Hily-Blant19} and \citet{Yoshida24}.  This is further traced when comparing the $^{12}$C/$^{13}$C as estimated for CO and CCH.  In TW Hya  CCH emits in a ring at radii of $\sim$40-110~au \citep[Fig.~\ref{fig:CCH_zeroth};][]{Bergin16} and our detection of \ce{^{13}CCH} must arise from within the ring where we measure a ratio of $^{12}$C/$^{13}$C in CCH of 65$\pm$15.  In contrast, over a similar spatial extent, \citet{Yoshida22} estimate a $^{12}$C/$^{13}$C ratio of 21$\pm$5 in CO. If the isotopic ratio of 21 existed in CCH in our data then we would have a 14$\sigma$ detection $^{13}$CCH in the integrated emission, which is in contrast to the current 4$\sigma$ detection.
This re-enforces the evidence that there are two isotopic reservoirs for carbon present in TW Hya as argued previously by \citet{Yoshida22} for carbon.  We note two isotopic reservoirs have  also been isolated for nitrogen by \citet{Hily-Blant19}.

\begin{deluxetable*}{cclc}
\tablecolumns{4}
\tablewidth{0pt}
\tablecaption{ $^{12}$C/$^{13}$C in TW Hya \label{tab:tw_iso}}
\tablehead{
\colhead{Location (au)} \vspace{-0.2cm} & \colhead{Molecular Carrier} &\colhead{$^{12}$C/$^{13}$C Ratio} & \colhead{Reference}\\ \vspace{-0.2cm}}
\startdata
  0 -- 20  & \ce{^{12}C^{18}O}/\ce{^{13}C^{18}O} & 40$^{+9}_{-6}$ & \citet{Zhang17} \\  
    $<$\,60  & \ce{H^{12}CN}/\ce{H^{13}CN} & 86$^{+4}_{-4}$ & \citet{Hily-Blant19} \\
  25 -- 55 &  \ce{^{12}CN}/\ce{^{13}CN} & 70$^{+8}_{-6}$ & \citet{Yoshida24}\\
 70 -- 110  & \ce{^{12}CO}/\ce{^{13}CO} & 21$^{+5}_{-5}$ & \citet{Yoshida22} \\  
 40 -- 110  & \ce{^{12}CCH}/\ce{^{13}CCH} & 65$^{+20}_{-20}$ & This Paper \\  
 $>$\,130  & \ce{^{12}CO}/\ce{^{13}CO} & $>$\,84 & \citet{Yoshida22} \\  
\enddata
\vspace{-0.8cm}
\end{deluxetable*}

\section{Model of Disk Carbon Isotopic Chemistry}

To examine the origins of two isotopic reservoirs for carbon in the TW Hya disk we use a detailed model of the carbon isotopic chemistry for TW Hya. 
For this purpose, we use the Dust And LInes thermochemical model (DALI) code \citep{Bruderer12, Bruderer13}.  Our aim is to explore a generic model of the chemistry of carbon isotopic fractionation in two frameworks.  One is an extant model of the TW Hya disk that has been used in previous modeling efforts \citep{Trapman17, Bosman19_twhya}.  We call this model the ``late-stage disk'' model as it is motivated by the TW Hya Class II disk as it is today with numerous observational constraints.   In addition, we present an ``early-stage'' model that explores  the origins of isotopic chemistry during earlier phases more consistent with a Class I disk.  In our efforts we will not fit the overall emission levels; rather, we will explore whether two separate carbon isotopic reservoirs are created.  That is, we are not attempting to fully match this system but rather attempting to explore the conditions that may lead to carbon isotopic fractionation.

We use the extant TW Hya model published by \citet{Trapman17} and \citet{Bosman19_twhya}.  
The basic parameters of the model are given in the appendix.  The chemical network including isotopic fractionation is created by \citet{Miotello14} and includes key fractionation reactions linking $^{12}$C and $^{13}$C \citep{Langer84, rollig13}.  The critical reactions in this case are the primary fractionation reaction:
\begin{equation}
{\rm ^{13}C^+ + CO \leftrightarrow C^+ + ^{13}CO + \Delta E = 35~K}, 
\end{equation}

\noindent which favors $^{13}$CO at low ($\lesssim$40~K) temperature.   The other key isotopic fractionation pathway in the disk is isotopic selective photodissociation of CO and we use the rate prescription of \citet{visser09}.   For a basic introduction to carbon isotopic fractionation in the disk framework we refer the reader to \citet{Woods09} and to the review by \citet{Nomura22}.    In this model we include a grain surface chemistry that is limited to simple hydrogenation reactions to make saturated species (e.g. water, ammonia, methane).  Further, the overall network does not extend to methanol.

\subsection{Late-Stage Model}
\label{sec:olddisk}

Here, we use the existing model of TW Hya by \citet{Trapman17} and \citet{Bosman19_twhya}.
This model has been designed to match several well characterized aspects of this disk.  First, the strong mm-continuum image and scattered light surface requires grain growth and settling \citep{vanBoekel17}. Further, the mm-continuum emission shows that the large grains are more spatially confined compared to the gas \citep{Andrews16, Huang18}.    We
 therefore, assume the disk is several Myr-old \citep{Sokal18} with 99\% of the grain mass confined to the midplane to match the mm-wave dust emission distribution.
 
  The primary ionization sources towards this disk (ultraviolet, X-ray, cosmic rays) have been constrained via observations.   For thoroughness the UV field is the stellar spectrum of TW Hya with UV excess from \citet{Cleeves15}. We adopt L$_X$ = 1.4 $\times$ 10$^{30}$~ergs~s$^{-1}$ with a plasma temperature of 3 MK \citep{Stelzer04}.  The cosmic ray ionization rate ($\zeta_{cr}$) is assumed to be 10$^{-19}$~s$^{-1}$ which is constrained via chemical analysis of molecular ion emission \citep{Cleeves15}.  Additional information is found in the original reference for this model \citep{Trapman17}.
 
  Analysis of numerous disk CO isotopologue observations, starting with TW Hya \citep{favre13a}, have found that the abundance of CO appears to be reduced in disk gas \citep[see discussion in][]{Miotello_ppvii}.  Essentially, CO is believed to be provided to the disk with an ``interstellar'' abundance relative to H$_2$ $\sim$ 10$^{-4}$ \citep{Bergin17}.   Some regions of the disk have temperatures below the CO sublimation temperature ($\sim$20~K) and CO ice forms.  However, in regions where the dust temperature is $>$ 20~K, the expectation is that the CO gaseous abundance will be interstellar.  However, in systems where strong constraints on the H$_2$ mass exist, such as TW Hya, the CO gas phase abundance is found to be depleted, i.e. [C/H]$_{gas}$ $<$ [C/H]$_\star$ \citep{favre13a, Kama16b, Zhang17, Zhang19}. This has now been confirmed via a direct measurement of the H$_2$ density through pressure broadened line wings \citep{Yoshida22_linebroad}.  The depletion of CO \citep[and also water vapor;][]{Du17} is believed to occur via chemical processing of CO into species with lower volatility, such as CO$_2$ or CH$_3$OH, and via interactions with grain evolution \citep[see][for comprehensive model]{Krijt20}.
In sum, the \citet{Trapman17} model assumes that overall gas-phase carbon is depleted relative to ISM levels, with the missing carbon present as ices on the mm-grains that are settled to the midplane and more spatially confined than the gas.

\begin{table}
\centering
\caption{\label{tab:chemabun}Initial Chemical Abundances$^{a}$}
\begin{tabular}{rll}
\hline \hline
\multicolumn{1}{c}{Species} & \multicolumn{1}{c}{Late Stage} & \multicolumn{1}{c}{Early Stage}  \\
\hline
H$_2$                & 0.5  & 0.5\\
He               & 7.6 $\times$ 10$^{-2}$ & 7.6 $\times$ 10$^{-2}$\\
Mg              & 4.2 $\times$ 10$^{-10}$  & 4.2 $\times$ 10$^{-10}$\\
Si               & 7.9 $\times$ 10$^{-9}$ & 7.9 $\times$ 10$^{-9}$ \\
S               & 1.9 $\times$ 10$^{-9}$  & 1.9 $\times$ 10$^{-9}$\\
Fe               & 4.3 $\times$ 10$^{-10}$    & 4.3 $\times$ 10$^{-10}$  \\
CH$_4$             & 2.7 $\times$ 10$^{-6}$ & \nodata \\
$^{13}$CH$_4$            & 3.5 $\times$ 10$^{-8}$ & \nodata \\
CO               & 2.7 $\times$ 10$^{-6}$    & 2.7 $\times$ 10$^{-4}$\\
$^{13}$CO           & 3.5 $\times$ 10$^{-8}$ & 3.5 $\times$ 10$^{-6}$\\
N$_2$              & 1.1 $\times$ 10$^{-5}$ & 1.1 $\times$ 10$^{-5}$\\
N$^{15}$N           & 2.4 $\times$ 10$^{-8}$ & 2.4 $\times$ 10$^{-8}$\\
H$_2$O$_{ice}$ & 2.6 $\times$ 10$^{-6}$ & 2.6 $\times$ 10$^{-6}$\\
\hline
\multicolumn{3}{l}{$^a$Relative to H}\\
\end{tabular}
\end{table}

 Finally, we assume C/O $\sim$2 based on the analysis from C$_2$H \citep{Bergin16, Kama16a}.
 This model assumes that 50\% of the remaining gaseous carbon is in CO and 50\% originally in CH$_4$.    The presence of the CH$_4$ has been previously shown to generate strong C$_2$H
emission in disks \citep{Bosman21}.\footnote{Specifically this paper states that the origin of C$_2$H arises from CH$_4$ released from carbon in refractory organics.  However, this is simulated in the system by putting carbon in CH$_4$.}  Our model is run for 1 Myr, which is much less than the estimated age of TW Hya \citep[$\sim$8~Myr;][]{Sokal18}.   This difference is because we are not modeling the CO abundance evolution, which we assume took place during earlier stages.  Rather our goal is to explore whether existing TW Hya chemical models can readily fractionate carbon carriers to create two interdependent and stable isotopic reservoirs.    Table~2 provides the overall initial chemical abundances.

\begin{figure*}
    \begin{center}
    \includegraphics[width=1.0\textwidth]{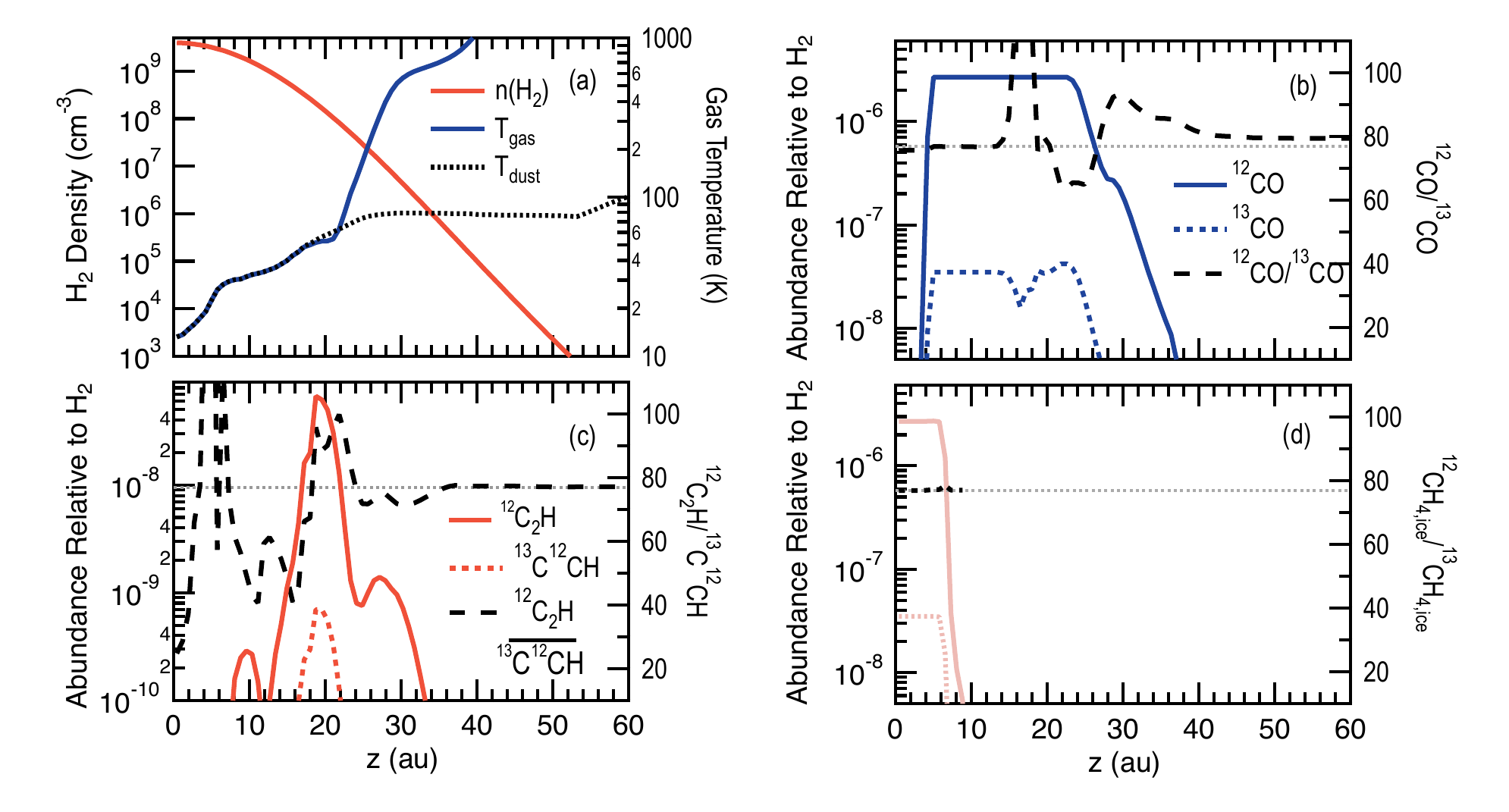}
    \caption{ Predicted vertical distribution of CO, CCH, and CH$_4$ ice isotopic ratios in the ``late-stage disk'' model at a radial location commensurate with the center of the C$_2$H emission ring (63 au). (a) Overall gas density structure and modeled vertical gas thermal profile at this radius.
    (b,c,d) Predicted vertical distribution of {\rm CO/$^{13}$CO}, {\rm C$_2$H/$^{13}$CCH}, and CH$_{\rm 4,ice}$/$^{13}$CH$_{\rm 4,ice}$ from the thermochemical model.  In these panels the chemical abundances are referenced to the left axis while the dashed line provides the predicted isotopic ratio that is referenced to the axis on the right. For all species the initial isotopic ratio is 77 which is shown as the dotted grey line.  }
    \label{fig:olddisk}
    \end{center}
\end{figure*}

\begin{figure*}
    \centering
    \includegraphics[width=0.8\textwidth]{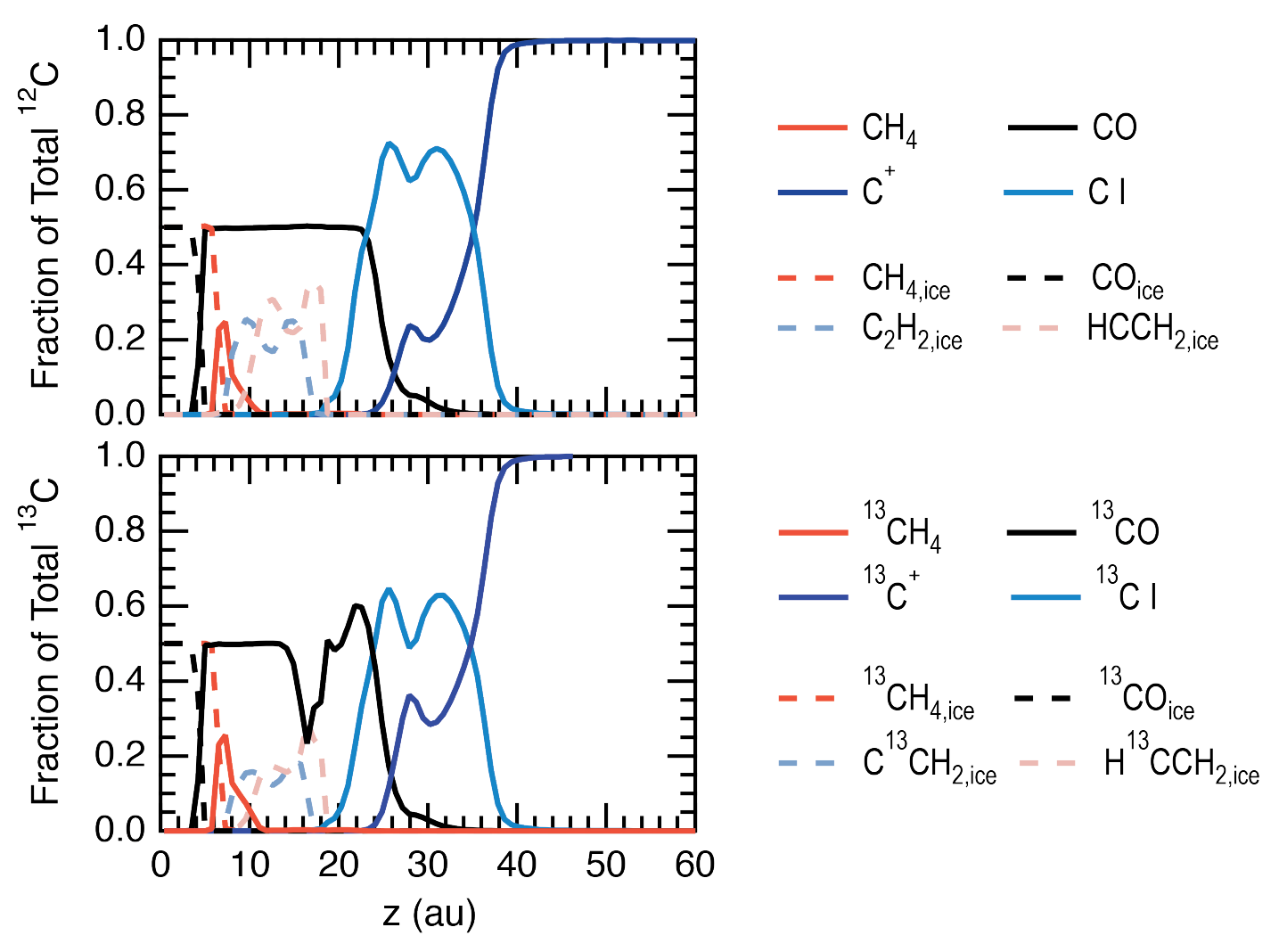}
    \caption{Plot of (top) major carriers of $^{12}$C and (bottom) $^{13}$C.  Abundances are normalized to the total amount of isotopic carbon with $x$($^{12}$C) = $5.2 \times 10^{-6}$ and  $x$($^{13}$C) = $7.0 \times 10^{-8}$ (abundances given relative to total H). This model is sampled at r=63 au, the location of the C$_2$H emission ring.   We note that HC$^{13}$CH$_2$ is not shown here but has the same distribution and abundance as H$^{13}$CCH$_2$. 
       }
    \label{fig:old-full}
\end{figure*}

With this calibrated model we explored whether this physical perspective could produce two separate isotopic reservoirs.  For this purpose, we include a network that includes fractionation of 
carbon \citep{Miotello14}. 
The results are shown in Fig.~\ref{fig:olddisk}.
In this Figure, panel (a) provides the vertical density and thermal profile. 
In panels (b-d) we show model predictions as a vertical cut 
taken at 63~au.  This location traces the C$_2$H ring and is commensurate with where C$_2$H and CO appear to trace two separate isotopic reservoirs (Fig.~\ref{fig:isointwhya}).  The figure shows the C$_2$H, CO, and CH$_4$  vertical distributions in abundance alongside the $^{13}$C isotopologues.  In our models the initial isotopic ratio in both CO and CH$_4$ is 77 and we seek to determine if the overall isotopic chemistry produces two separate reservoirs. 
The full chemical evolution from this model with the major carriers shown is given in Fig.~\ref{fig:old-full}. 

Exploring panel (b) in Fig.~\ref{fig:olddisk} we find an interesting fractionation pattern for CO.  At the upper surface, near z = 25~au, the effects of CO isotopic selective photodissociation are found with a deficit in $^{13}$CO.   Deeper in the disk there is a dip in the $^{13}$CO abundance near z = 17~au that produces a sharp increase in $^{12}$CO/$^{13}$CO over a short vertical distance.    The gaseous CO isotopic ratio estimated from column density ratios at 63 au is 79 which is well above the observed value of 21.

 In layers above the midplane (4~au $<$ z $<$ 20~au) where CH$_4$ ice sublimates, much of the excess carbon provided by CH$_4$ is processed into hydrocarbon ices (C$_2$H$_2$ and HCCH$_2$; Fig.~\ref{fig:old-full}).    
Above the edge of this vertical zone near 20~au gaseous C$_2$H production peaks.  For C$_2$H, the model predicts a complex fractionation pattern.  However, it is important to focus on vertical layers where C$_2$H is found in abundance between 14~au $<$ z $<$ 24~au.   Here, some of the excess carbon resulting from isotopic selective photodissociation of CO finds its way into C$_2$H raising the $^{13}$CCH abundance above the initial value, with a sharp decrease in CCH/$^{13}$CCH at z = 18~au as the $^{13}$C finds its way into hydrocarbon ices.   The gaseous C$_2$H isotopic ratio estimated from column density ratios is 78 which agrees with observations (Table~1).  The predicted total C$_2$H column density at this radius is 6 $\times$ 10$^{14}$~cm$^{-2}$.  This would give an integrated flux density for the N=4--3 J=9/2--7/2 F-5-4 transition of $\sim$0.2 Jy km/s in a 0.45$''$ beam at 60 pc (assuming a gas temperature of 40~K).  This is close to the observed peak integrated flux density within the emission ring (see Fig.~\ref{fig:CCH_zeroth}).  We also confirm that the gas phase C/O ratio in the model is 2  in the C$_2$H emitting layer, which is consistent with previous work \citep{Bergin16}.

In summary, there is a narrow range of physical parameters where $^{13}$C fractionation is found, which limits the impact over the entire column.    This particular ``late-stage disk'' model is incapable of producing two separate isotopic ratios for CO and C$_2$H (and HCN).    As a final check, to explore whether the excess CH$_4$ in the gas could hide the presence of isotopic fractionation, we ran an identical model with no excess CH$_4$ as an initial condition.  This model did not produce significant carbon isotopic fractionation of CO or C$_2$H.  For this set of initial conditions, the C/O ratio is effectively solar at these distances (i.e. C/O = 1) and, as a result, the C$_2$H column was reduced by 3 orders of magnitude.  This is consistent with earlier analyses \citep{Bergin16, Cleeves18, Bosman21_mapsco}.

\subsection{Early-Stage Disk Model}

\subsubsection{Motivation}

Our model of TW Hya as it exists today is incapable of producing two separate carbon isotopic reservoirs.  We therefore also explore a model of a younger disk where conditions might be significantly different.
  Our hypothesis is as follows.  Numerous works have shown that, if cosmic rays are present, CO can be chemically processed into a variety of less volatile species which may relate to the reduced abundance of CO as seen in numerous disk systems \citep{Bergin14, Furuya14, Reboussin15, Eistrup17, Schwarz18, Bosman18}.   Perhaps this process might also lead to isotopic fractionation of CO and thereby simultaneously set the stage where another carbon-bearing molecule would carry the opposite isotopic signature.  If this carbon-bearing molecule were a hydrocarbon then it might provide the excess carbon needed to generate strong C$_2$H emission at later stages.

   The chemical processing of CO appears to be time dependent as two independent analyses of C$^{18}$O and lesser abundant isotopologues  find a decline in the gaseous CO abundance from the initial interstellar level as a function of evolutionary state corresponding to $\sim$1 Myr timescales \citep{Bergner20, Zhang20}.   Along these lines the gaseous C/O ratio might also change as the disk evolves.  Interstellar ices have some CO and CH$_4$ \citep{oberg11_c2d}.  These ices would sublimate at 30~K \citep{Minissale22} to mix with CO gas that has C/O = 1.  However, interstellar CH$_4$ ice is only $\sim$10\% of the CO ice content \citep{McClure23}.   When both sublimate to the gas at 30-40~K (with water and CO$_2$ frozen), the resulting composition is effectively C/O $\sim$ 1.  Given that observational analyses are suggesting that TW Hya has C/O $\sim$ 1.5-2 there also must be some additional chemical evolution in the volatile inventory and its C/O ratio.
  
  In this light, cosmic rays are known to be present in well characterized pre-stellar cores with measured ionization rates consistent with expectations for the dense interstellar medium \citep{Maret07, Redaelli21}.  In contrast,
  Class II systems appear to have reduced ionization that are argued to be the result of cosmic ray exclusion by magnetized winds (analogous to the solar system's heliosphere) or tangled magnetic fields within the disk environment \citep{Cleeves15, Aikawa21, Seifert21}.    
   Class I systems still have residual envelopes that limit the reach of magnetized stellar winds, restricting their effect to the extent of the cavity carved by the protostar's energetic flows 
 \citep{Arce07, Hseih23}.   Within the outflow cavity it is possible that cosmic rays are excluded.  However,  
energetic particles might still penetrate through envelope and into the protostellar disk through equatorial zones.  Overall, it is possible that changes in the ionization rate are associated with the clearing of the envelope from the protostellar to the protoplanetary disk stage. 
It has long been recognized that cosmic ray ionization is central towards the isotopic fractionation of carbon in the interstellar medium \citep{Langer84, Langer89, Furuya2011, Roueff2015}.  Perhaps ionization evolution might also provide some imprint on isotopic ratios carried through subsequent evolutionary stages.

\subsubsection{Model and Results}

  To model an earlier stage TW Hya {\em protostellar } disk  (e.g., Class 0/I),
 we assume the grains have grown but only 90\% of their mass is in large grains (i.e., mm-sized).  We also include the presence of cosmic rays ($\zeta_{\rm H_2} = 5 \times 10^{-17}$~s$^{-1}$).  This value is at a level a factor of 50 above that currently estimated towards TW Hya by \citet{Cleeves15} which was used in the late-stage disk simulation.  The initial chemical abundances, given in Table~\ref{tab:chemabun}, place the initial CO abundance at 2.7 $\times$ 10$^{-4}$ (i.e. interstellar) with no additional carbon in any form. Thus, we assume that there has been no chemical processing of CO as assumed in the earlier model. 
The initial CO isotopic ratio is also set to an  interstellar isotopic ratio of 77.  At the start of the simulation the only reservoir of $^{12}$C and $^{13}$C is CO. 
 This model has some water ice included with an abundance 2.7 $\times$ 10$^{-6}$ and is run for 1 Myr. Here we assume that a significant amount of water ice has accumulated in the midplane where most of the dust mass still resides \citep{Krijt16}.
 The baseline model structure parameters are given in the appendix.

\begin{figure*}
    \centering
    \includegraphics[width=1.0\textwidth]{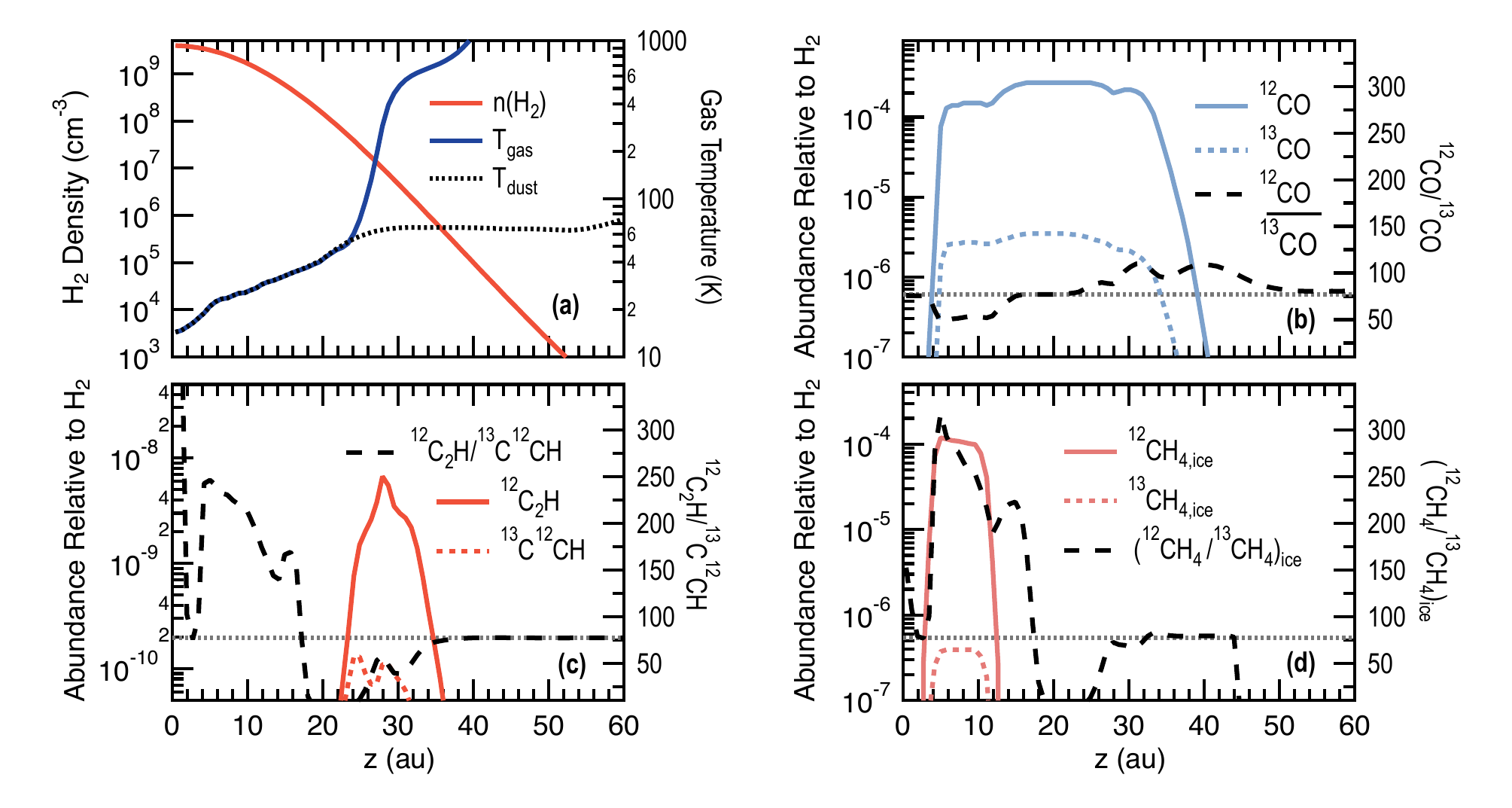}
    \caption{Predicted vertical distribution of CO and CCH isotopic ratios in the ``early-stage disk'' model at a radial location commensurate with the center of the C$_2$H emission ring (63 au). (a) Overall gas density structure and modeled vertical gas thermal profile at this radius.
    (b), (c), (d) Predicted vertical distribution of {\rm $^{12}$CO/$^{13}$CO}, $^{12}$C$_2$H/$^{13}$C$^{12}$CH, and {\rm $^{12}$CH$_4$/$^{13}$CH$_4$} from the thermochemical model.  In these three panels the chemical abundances are referenced to the left axis while the dashed line provides the predicted isotopic ratio that is referenced to the axis on the right. In this model all $^{13}$C is placed in CO at the start with an initial isotopic ratio of 77, which is shown as the dotted grey line. This model is sampled at r=63 au, the location of the C$_2$H emission ring.
       }
    \label{fig:young}
\end{figure*}

\begin{figure*}
    \centering
    \includegraphics[width=0.8\textwidth]{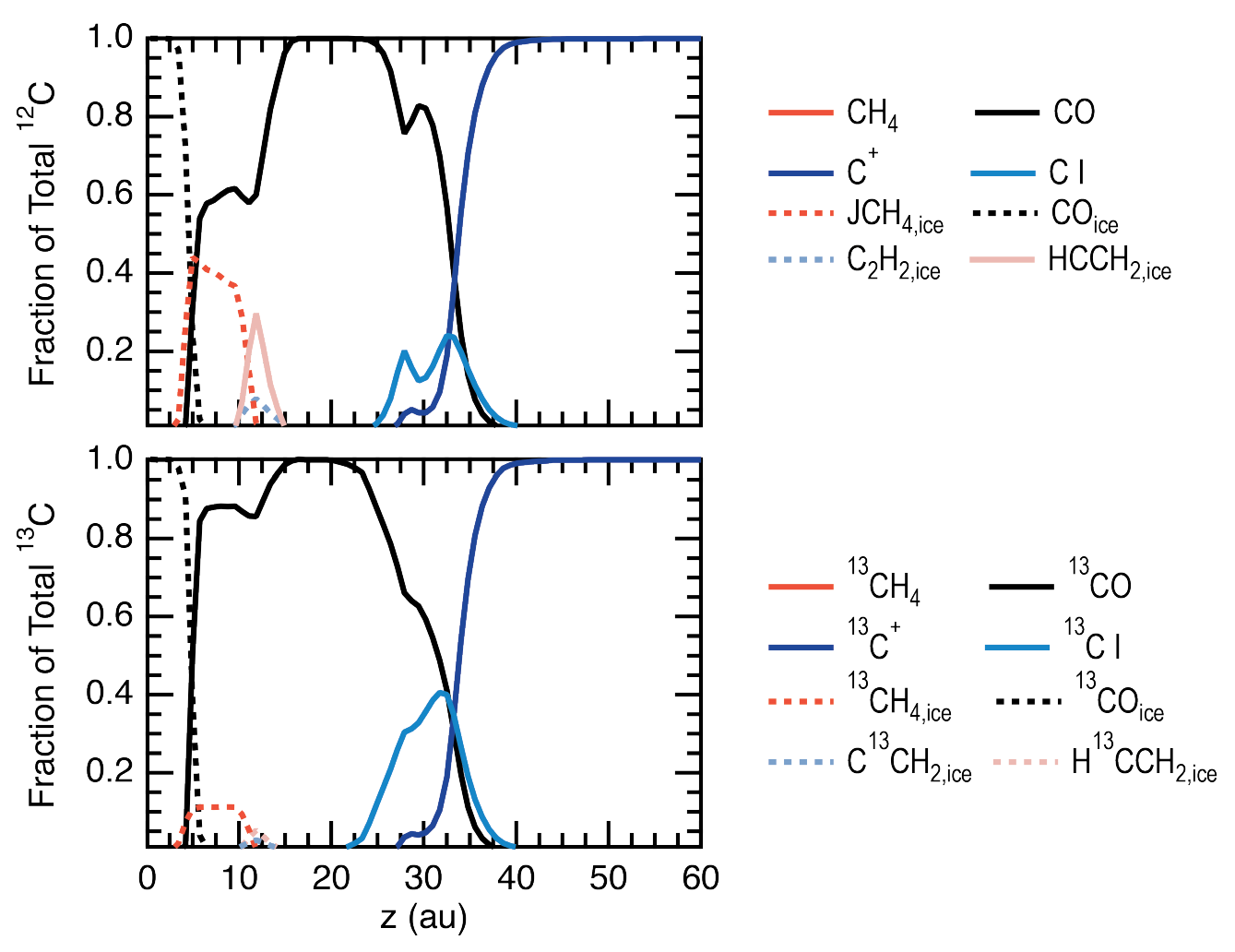}
    \caption{Plot of (top) major carriers of $^{12}$C and (bottom) $^{13}$C.  Abundances are normalized to the total amount of isotopic carbon with $x$($^{12}$C) = $2.7 \times 10^{-4}$ and  $x$($^{13}$C) = $3.5 \times 10^{-6}$ (abundances given relative to total H). This model is sampled at r=63 au, the location of the C$_2$H emission ring.
       }
    \label{fig:young-full}
\end{figure*}

 Fig.~\ref{fig:young} presents the results from this model at r = 63~au in the form of vertical cuts of the abundance distribution of key isotope carriers.  Fig.~\ref{fig:young-full} provides the overall carbon inventory over this same region.
 Comparison of the thermal structure between the late-stage (Fig.~\ref{fig:olddisk}) and early-stage (Fig.~\ref{fig:young}) disk model show that the less evolved system, with a larger mass in small grains, traps radiation higher in the disk leading towards slightly cooler deeper layers.  Further the gas phase abundance of CO is distributed within the warm molecular layer \citep[i.e., above the CO freeze-out layer in the midplane and below the photodissociation layer;][]{Aikawa02} which exists over a wider range in vertical distance (Fig.~\ref{fig:young}b).
 As in the late-stage disk model (Fig.~\ref{fig:olddisk}), C$_2$H forms at intermediate heights albeit in slightly higher layers in the early-stage disk model shown in panel (c) in Fig.~\ref{fig:young}.
 One additional difference for C$_2$H is the peak abundance of C$_2$H is reduced by about an order of magnitude as the amount of free carbon is reduced in this model.  The total C$_2$H column density from this model (at r = 63~au) is 5 $\times$ 10$^{12}$ cm$^{-2}$.  This is two orders of magnitude reduced from the late-stage model. In this model the C/O ratio in the gas is near unity and a reduction in the C$_2$H column/line flux density is expected.

In the early-stage disk model at high altitude (z=35~au; Fig.~\ref{fig:young}b) CO is  destroyed via photodissociation.
Within the  layer where CO is photodestroyed the effects of isotopic selective photodissociation of CO are observed and $^{12}$CO/$^{13}$CO exceeds the assumed initial value shown as the dotted line.   
More substantive changes are seen deep within the warm molecular layer where cosmic rays chemically process about 40\% of gas phase CO into methane ice (Fig.~\ref{fig:young}b and c). 
 The chemical processing of CO is initiated via reactions with He$^+$ atoms which free $^{12}$C$^+$ and $^{13}$C$^+$ alongside O atoms \citep{aikawa96}.  The oxygen atoms collide with cold dust grains and hydrogenate to form water ice. In this model about 40\% of the O in CO is placed into water ice leaving excess C$^+$ in the gas.  This causes the drop in the CO abundance near z = 14~au Fig.~\ref{fig:young}b).  

 This excess ionized carbon would normally find its way to CO, but the lack of oxygen means it sits in a sea of H$_2$ gas.  As noted by \citet{Du15} this leads to the formation of unsaturated and saturated hydrocarbons that work towards species that can freeze onto grains when their individual sublimation temperature is above the temperature of the dust.  In this instance the {\em gas-phase} chemistry favors the formation of mostly  \ce{CH4} ice but also  \ce{C2H2} and \ce{HC2H2} ices.

In the layer where gas phase CO chemical processing occurs, cosmic rays are also powering carbon isotopic fractionation.
 In this case, the He$^{+}$ destruction of CO and $^{13}$CO produces an excess $^{13}$C$^+$ atoms which are favored to be implanted into $^{13}$CO via reaction (1) due to the low gas temperature in these layers (T$_{gas} <$ 40~K).  In all, $^{13}$CO gas becomes modestly enriched with $^{12}$C/$^{13}$C $\sim$ 50 compared to the initial value of 77.
Because the $^{13}$C becomes locked in gaseous CO, the hydrocarbon ices carriers have $^{12}$C/$^{13}$C $>$ ($^{12}$C/$^{13}$C)$_{\rm ISM}$.

\begin{figure}
    \centering
    \includegraphics[width=0.5\textwidth]{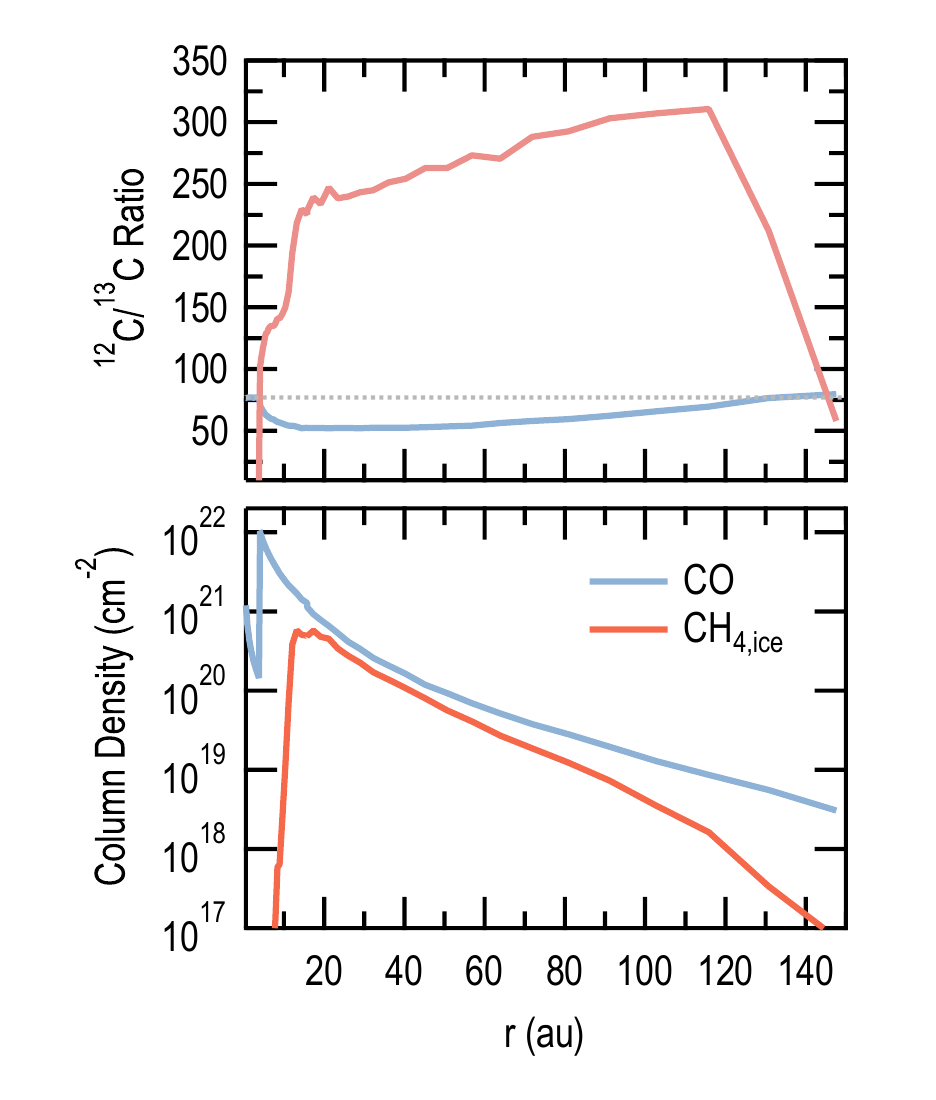}
    \caption{(top) predicted \RCO\ in CO and \ce{CH4} ice as a function of radius from the early-stage disk model.  Isotopic ratios are derived through the ratio of the total column densities of the respective isotopologue. (bottom) Predicted total column density profiles for gas phase \twCO\ and solid state \ce{CH4} as a function of radius for both molecules.  Inside $\sim 12$~au the CH$_4$ iceline is reached in the midplane and other more complex hydrocarbons with assumed higher sublimation temperatures (e.g. HCCH$_2$) are created.  The drop in the gas phase CO abundance near 4~au is due to an assumed inner dust cavity.
       }
    \label{fig:youngvsR}
\end{figure}

   Similar behavior is seen throughout the disk, but at variable height about the midplane.  In Fig.~\ref{fig:youngvsR} we provide the $^{12}$C/$^{13}$C ratio in CO and methane ice as a function of radius.  For this plot we have computed the ratio from the estimated vertical column densities.  This demonstrates that this model effectively produces two isotopic reservoirs for carbon across much of the disk.
   In the bottom panel of the Figure, we also provide the column densities of the main isotopologues.  The comparable column density between gaseous CO and \ce{CH4} ice shows that this effect is present both radially and vertically.  We note that the midplane (z $<$ 3 au) remains unchanged by this evolution with CO ice as the dominant carbon carrier for both isotopologues and the ratio fixed at 77.

   One issue that we have not addressed is fact that the CO isotopic ratio appears to be elevated above ISM values beyond 130~au (see Fig.~\ref{fig:isointwhya}).  Earlier work by \citet{Zhang19} inferred extreme CO depletion values beyond 100 au (i.e., abundances reduced by factors of 100 more compared to interstellar levels).  They suggested that this may be the result of gaseous dissipation of the disk. If this is the case, then the elevated CO isotopic ratio at these large distances could be the result of isotopic selective photodissociation.  If this is the case, then perhaps isotopic ratios might be used as a probe of gas disk dissipation. 
  
\subsubsection{Critical Dependencies}

 We have made a number of assumptions in this model including the presence of cosmic rays, increasing the amount of small grains, and reducing the amount of water ice present in upper layers.   Of these we find that the first two are important for the generation of the isotopic enrichments.  Cosmic ray ionization has the highest penetration power, and if present, clearly will dominate the ion-molecule chemistry leading to fractionation.  However, high levels of hard X-rays ($>$ 1 keV), with scattering, might also be important \citep{Igea99}.  Increasing the ratio of small to large grains (i.e. more mass in small grains) leads to a more flared surface and a higher mass present below 35~K (e.g. compare Figs~\ref{fig:olddisk}a and Fig.~\ref{fig:young}a).  In this model this effect is not large ($\sim$20\% more mass below 35~K in the early-stage model), but can be expected to be higher with less dust evolution and higher flaring angles.  Another aspect is the greater amount of small grains increases the effective surface area ($<n\sigma>_{gr}$) of grains, enhancing the dust/gas interaction and enabling greater freeze-out of methane.   For the last factor, the assumed water ice abundance, we find that if we assume an initial water ice abundance in the early-stage model of 10$^{-4}$ (relative to H) as opposed to the original value of 2.6 $\times$ 10$^{-6}$ then there is no major difference in the overall chemistry compared to the results presented in Figs.~\ref{fig:young} and \ref{fig:young-full}.  While fractionation levels and abundances of key carriers (CO, CH$_4$) are unaffected, the C$_2$H column is altered and is reduced by an order of magnitude.

\section{Discussion}

For the past decade observational analyses of resolved gas-rich disks have revealed two puzzles.  One regards the overall abundance of CO which appears to be below the expected value of 10$^{-4}$ relative to \ce{H2} \citep{favre13a, Schwarz16, Bergin17, Miotello17, Miotello_ppvii}.  The other is the strong emission from \ce{C2H} which rivals that of $^{13}$CO \citep{Kastner14, Bergin16, Miotello19} and has been attributed to a C/O ratio that exceeds unity.  In the latter case, we cannot rule out a source term for the excess CO in TW Hya from the photoablation of refractory carbon grains as discussed by \citet{Bosman21}.
However, the detection of  enriched $^{13}$CO within an exoplanetary atmosphere \citep[][]{Zhang21_13coexo} requires a source term for the enrichment.  That source term is apparently not carbon isotopic fractionation in the cold interstellar medium based on initial results of ice abundances from JWST \citep{McClure23}.   

In this initial exploration, the evolution of  an early-stage disk exposed to cosmic ray ionization offers a potential avenue to generate a $^{13}$C enrichment for CO.  It simultaneously provides a source term for gas phase hydrocarbons in evolved disk systems whose chemical formation pathways can potentially be powered by photodesorption of hydrocarbon ices. \citet{Bosman21} shows that if this carbon is released into the gas via photodesorption, it can generate a long-lived photochemical equilibrium cycle.

\subsection{Disk Ionization}
The central need to generate the carbon isotopic fractionation points to the presence of an ionization source deep in the disk. To reach significant mass this source of ionization likely needs the penetrating power of interstellar cosmic rays \citep{Eistrup17, Schwarz18, Bosman18}.   Based on an analogy to the Sun, and detailed models of cosmic ray-stellar wind interactions it is thought that the interstellar cosmic rays responsible for ionization of H$_2$ are not present within  disk systems that have dissipated their natal envelopes \citep[e.g., class II systems;][]{cleeves13a, Fujii22}. Analysis of the observed molecular ion chemistry in class II systems also appear to support reduced interstellar cosmic ray ionization rates \citep{Cleeves15, Aikawa21, Seifert21}.  

However, the timescales of the CO abundance evolution in disk systems suggest a shift between the protostellar phase, where the envelope is present (e.g., Class 0 and I), and the protoplanetary disk phase \citep{Zhang20, Bergner20}.  Thus, chemical processing might occur via ionization during earlier evolutionary phases.   Indeed, numerous combined observational efforts,  over much larger spatial scales, suggest that particle ionization rates might be quite high ($>$10$^{-15}$~s$^{-1}$) during some portion of the protostellar phase \citep{Ceccarelli14, Favre18, Cabedo23, Lattanzi23}.  This is supported by theoretical work that suggest cosmic rays could be generated via shocks within the jets \citep{Padovani16, Gaches18}.   It is worth nothing that these observations are obtained on a spatial scale much larger than the young disk. Evidence for enhanced ionization rates is not universal as \citet{VantHoff22} measures a cosmic ray ionization rate below the ISM value in the L1527 protostellar disk.
However, if ionization rates are elevated, even for a short timescale, then chemical processing of CO within infalling material or in the protostellar disk might provide the needed ionization to generate  multiple isotopic reservoirs.

\subsection{Caveats and the Link to Planets}

From the perspective of planet formation and the detected $^{13}$C enrichment in an exoplanet by \citet{Zhang21_13coexo}, the combination of this work and \citet{Yoshida22}  characterizes  the carbon isotopic ratios within the main carbon reservoirs in gas with C/O $>$ 1.  In this gas the $^{13}$C carrier is gas-phase CO. 
If the early-stage disk composition  persists to later stages, as observed in TW Hya today, then perhaps gas phase carbon provides the $^{13}$C enrichment and not pebbles.  
One complication is that our early-stage disk model places the $^{13}$C deficit as ice coatings  of small grains in the upper layers (z = 5-15~au).  The evolution of these grains somehow must be separated from CO via dust evolution.  Since CO carries the opposite isotopic signature if both reservoirs were  available, they would effectively cancel out the enrichment when supplied to a planet.  How this might occur is uncertain and more detailed models of coupled gas/grain evolution are needed.  

Our models also do not fully match observations.  The level of $^{13}$CO enrichment in our early-stage disk model (assuming no additional) evolution is only $^{12}$CO/$^{13}$CO = 50 and not the observed level of 20 \citep{Yoshida22}.   Further, the level of $^{12}$C enrichment ($^{12}$C/$^{13}$C $\sim$ 250) in hydrocarbon ices, if they are source terms for C$_2$H in later stages, is a factor of 3 or more above the carbon isotopic ratio inferred from C$_2$H via the detection of $^{13}$CCH presented here.  
Our models suggest that, in the context of an assumed low cosmic ray ionization rate, little CO fractionation occurs in the dense disk.  Thus, over time the active chemistry in surface layers, perhaps with weak turbulent mixing, would tend to lower the $^{12}$C/$^{13}$C ratio in C$_2$H towards the value found for CO. Thus, time evolution could be an answer for C$_2$H. For CO the higher levels of $^{13}$C enrichment would require a different model that is colder and/or with higher ionization rates than the early-stage model adopted here.
Regardless the detection of two isotopic reservoirs in carbon \citep{Yoshida22} and nitrogen \citep{Hily-Blant19}
is providing information on the overall evolution of disk gas.

\subsection{Ionization Evolution: Model Predictions}

This work highlights the importance of sources of ionization in the dense portion of the disk. Our late-stage model, which matches other aspects of the TW Hya disk \citep{Trapman17, Bosman19_twhya}, assumes that the cosmic ray flux impinging on the disk is reduced  based on detailed analyses of molecular ion emission in the TW Hya disk \citep{Cleeves15}.  Under that assumption this model cannot create two disparate carbon isotopic reservoirs.  
In our work we highlight the potential for ionization evolution as a possible solution to some aspects of this conundrum.  This does lead to some testable predictions.   If the methane (or hydrocarbon-rich) ice created via CO chemical processing is the source term for the excess carbon that powers C$_2$H formation, then that ice is predicted to have an extreme carbon isotopic ratio.  Further, protostellar disks would have higher levels of methane/hydrocarbon ice (relative to CO ice) when compared to interstellar ices.  If this $^{13}$C poor material (with $^{12}$CH$_{4,ice}$/$^{13}$CH$_{4,ice}$ $\sim$150--300) is the source term for today’s TW Hya isotopic reservoir then some mixing, as discussed above, is required to lower the isotopic ratio.

Another possibility is that the destruction of refractory carbon powers the formation of C$_2$H as suggested by \citet{Bosman21}.   In the solar system, refractory carbon carries little carbon isotopic enrichment and is consistent with the interstellar medium \citep{Nomura22}.   This source term would be consistent with our late-stage model for C$_2$H, but this model cannot fractionate CO.   In this case detailed models will be needed to explore how the isotopic signature of this carbon would mix with that of CO to match observations.

\section{Summary}

We present the first detection of $^{13}$CCH in the TW Hya protoplanetary disk system and a detailed analysis of carbon isotopic fractionation in disk systems.  Our primary results are as follows.

\begin{enumerate}
    \item We determine an isotopic ratio for C$_2$H of CCH/$^{13}$CCH $= 65 \pm 20$ in gas sampling the same radii where \citet{Yoshida22} find CO/$^{13}$CO $= 21\pm5$. As C$_2$H is posited as the primary tracer of excess carbon in disks with C/O $\ge 1$, this confirms a carbon isotopic dichotomy in this system as suggested by \citet{Yoshida22}.  This is comparable to a similar result found for nitrogen by \citet{Hily-Blant19}.
\item We explore the origins of the carbon isotopic dichotomy through the use an extant thermochemical model of the disk \citep{Trapman17, Bosman19_twhya} with a full model of carbon isotope chemistry \citep{Miotello14} in a disk with C/O $>$ 1 with a reduced carbon content as consistent with observational constraints \citep{Bergin16, Kama16a, Bosman19_twhya, Yoshida22_linebroad}.   We find that this model is incapable of reproducing two independent reservoirs primarily due to the lack of cosmic ray ionization in deep disk layers where the temperature would be low enough to power the primary gas phase carbon isotopic fractionation reaction.  Additional efforts are needed to model the system adopting all available constraints on both the gas temperature \citep{Calahan21twhya} and ionization \citep{Cleeves15} simultaneously with the isotopic fractionation.

\item We also explore a model of the TW Hya disk in an earlier evolutionary state where the  initial C/O ratio is unity, interstellar cosmic rays are present, and there is no carbon depletion.    We find that this model can create two independent isotopic reservoirs one $^{12}$C-rich carried by methane and other hydrocarbon ices and the other $^{13}$C-rich carried by gas phase carbon monoxide. This dichotomy is predicted to be present throughout the disk and may lay the seeds for the future chemical evolution of the disk as CH$_4$ is a potential source term for the strong C$_2$H emission in disk systems \citep{Bosman21}.
\item We discuss the implications of this result for exoplanetary systems where gas phase CO appears to be the carrier of $^{13}$C enrichments.  Additional observational constraints regarding the carbon isotopic ratio within both Class I and Class II disks are sorely needed.
Further,  an understanding of the ionization evolution of disk systems is critical.

\end{enumerate}

\acknowledgements
We are grateful to the thorough referee who provided comments that improved this contribution.
 E.A.B. acknowledges support from NSF grant No. 1907653 and NASA NASA's Emerging Worlds Program, grant 80NSSC20K0333, and Exoplanets Research Program, grant 80NSSC20K0259.
KW's research was conducted at the Jet Propulsion
Laboratory, California Institute of Technology under contract with the National Aeronautics and Space Administration. KW was supported by a grant from NASA/Emerging Worlds program (18-EW182-0083). 
 L.I.C. acknowledges support from NASA ATP 80NSSC20K0529, the David and Lucille Packard Foundation, and the Research Corporation for Scientific Advancement Cottrell Scholar Award.

This paper makes use of the following ALMA data: ADS/JAO.ALMA\#2013.1.00198.S,

 \noindent ADS/JAO.ALMA\#2015.1.00308.S.
 
 \noindent ALMA is a partnership of ESO (representing its member states), NSF (USA) and NINS (Japan), together with NRC (Canada), MOST and ASIAA (Taiwan), and KASI (Republic of Korea), in cooperation with the Republic of Chile. The Joint ALMA Observatory is operated by ESO, AUI/NRAO and NAOJ. The National Radio Astronomy Observatory is a facility of the National Science Foundation operated under cooperative agreement by Associated Universities, Inc.
    \bibliography{z}

\appendix

\section{DALI model for TW Hya}

This model is based on the code and parameters as published by \citet{Trapman17} and modified by \citet{Bosman19_twhya}.   We adopt this model as a baseline and change a few parameters to investigate the chemical evolution at a slightly earlier evolutionary stage.

\begin{table}[!h]
\centering
\caption{\label{tab:model}TW Hya Model Parameters$^a$}
\begin{tabular}{l c}
\hline \hline
\multicolumn{2}{c}{Stellar parameters}\\
\hline
Stellar mass & 0.74 $M_\odot$\\ 
Stellar luminosity & 1 $L_\odot$\\ 
X-ray luminosity & $10^{30}$ erg s$^{-1}$ \\  
UV Luminosity & 2.7 $\times 10^{31}$ erg s$^{-1}$\\
\hline
\multicolumn{2}{c}{Disk parameters}\\
\hline
Disk mass & 0.025 $M_\odot$ \\
Critical radius ($R_c$) & 35 AU \\
Surface density slope ($\gamma$) & 1\\
scale-height at $R_c$ ($h_c$) & 0.1 rad\\
Flaring angle ($\psi$) & 0.3 \\
Inner radius & 0.05 AU\\
Gap inner radius & 0.3 AU \\
Gap outer radius & 2.4 AU \\
Inner disk gas-to-dust & 100 \\
Inner disk gas depletion & $10^{-2}$\\
Small dust size & 0.005--1 $\mu$m \\
Small dust fraction & 0.01 (late-stage), 0.10 (early-stage)\\
Large dust size & 0.005--1000 $\mu$m \\
Large dust fraction & 0.99 (late-stage), 0.90 (early-stage)\\
Large dust settling factor & 0.2 \\
Cosmic Ray Ionization Rate & 10$^{-19}$ s$^{-1}$ (late-stage), 5 $\times$ 10$^{-17}$ s$^{-1}$ (early-stage)\\
\hline
\multicolumn{2}{l}{$^a$ \citet{Trapman17, Bosman19_twhya}}\\
\end{tabular}
\end{table}

\end{document}